\documentclass[a4paper,11pt]{article}

\usepackage{contribution}

\newcommand{\weblink}[2][]{%
    \ifthenelse{\equal{#1}{}}%
    {\textnormal{\url{#2}}}%
    {\textnormal{\href{#2}{#1}}}%
}

\newcommand{\acknowledgements}[1]{%
  \bigskip\bigskip
  \textsf{\textbf{\Large Acknowledgements}} \\[2ex]
  {#1}
  \bigskip
}

\def\beq{\begin{equation}}
\def\eeq#1{\label{#1}\end{equation}}
\def\eeqn{\end{equation}}

\def\beqa{\begin{eqnarray}}
\def\eeqa#1{\label{#1}\end{eqnarray}}
\def\eeqan{\end{eqnarray}}

\let\bar=\overbar

\def\ie{{\it i.e.}}

\def\Dslash{\not{\hbox{\kern-4pt $D$}}}
\def\dslash{\not{\hbox{\kern-2pt $\del$}}}

\def\msb{{\bar{\ssstyle M \kern -1pt S}}}

\newcommand{\contribution}[7][]{%
  \clearpage
  \thispagestyle{plain}
  \ifthenelse{\equal{#1}{}}
  {\hypersetup{pdftitle={#2}}}
  {\hypersetup{pdftitle={#1}}}
  \hypersetup{pdfauthor={{#3} {#4}}}
  {\centering\normalfont\LARGE\bfseries\sffamily #2 \par\nobreak}
  \lhead{}
  \chead{%
    \textit{\footnotesize XIV International Conference on Hadron Spectroscopy
      (\weblink[\textit{hadron2011}]{http://www.hadron2011.de}), 13-17 June 2011, Munich, Germany}%
  }
  \rhead{}
  \bigskip
  \begin{center}
    {#3} {#4}\ifthenelse{\equal{#6}{}}{}{\footnote{\weblink[#6]{mailto:#6}}}
    \ifthenelse{\equal{#7}{}}{}{#7} \\
    \textit{#5}
  \end{center}
  \bigskip
}

\renewcommand{\abstract}[1]{%
  \begin{center}
    \begin{minipage}{0.85\textwidth}
      \begin{footnotesize}
        #1
      \end{footnotesize}
    \end{minipage}
  \end{center}
  \bigskip
}

\begin{document}

%
%
%
{  

\makeatletter
\@ifundefined{c@affiliation}%
{\newcounter{affiliation}}{}%
\makeatother
\newcommand{\affiliation}[2][]{\setcounter{affiliation}{#2}%
  \ensuremath{{^{\alph{affiliation}}}\text{#1}}}
%

\contribution[Spin-Flavor vdW and NN interaction]
{Spin-Flavor van der Waals Forces and NN interaction}
{A.}{Calle Cordon}  
{\affiliation[Thomas Jefferson National Accelerator Facility, Newport News, Virginia 23606, USA.]{1} \\
 \affiliation[Departamento de F\'isica At\'omica, Molecular y Nuclear, Universidad de Granada, E-18071 Granada, Spain.]{2}
}
{cordon@jlab.org}
{\!\!$^,\affiliation{1}$, E. Ruiz Arriola\affiliation{2}}
%

\abstract{%
\rule{0ex}{3ex}
We study the Nucleon-Nucleon interaction in the Born-Oppenheimer
approximation at second order in perturbation theory including the
$\Delta$ resonance as an intermediate state.
The potential resembles strongly chiral potentials computed either via 
soliton models or chiral perturbation theory and  
has a van der Waals like singularity at
short distances which is handled by means of renormalization
techniques. Results for the deuteron are discussed. 
%
%
%
%
 }
%

\section{Introduction}

A major goal of Nuclear Physics is the derivation of the 
Nucleon-Nucleon (NN) interaction from Quantum Chromodynamics (QCD). 
In QCD the fundamental degrees of freedom are colored quarks and
gluons which are \textit{confined} to form colorless strongly
interacting hadrons. 
Because of this the resulting nuclear forces at sufficiently large
distances correspond to spin-flavor excitations, very much like the
dipole excitations generating the van der Waals (vdW) forces acting
between atoms (for a review see e.g.~\cite{Feinberg:1989ps}). In the
Born-Oppenheimer (BO) approximation and assuming no retardation and no
electron cloud overlap at large distances, the atom-atom energy at a
separation distance $r$ can be calculated at second order perturbation
theory as,
\beqa
V_{AA} = \langle AA | V_{\rm dip} | A A \rangle +
 \sum_{AA \neq A^* A^*}\frac{|\langle AA | V_{\rm dip} | A^* A^* \rangle|^2}{E_{AA}-E_{A^*A^*}} + \dots  = -\frac{C_6}{r^6} + \dots \, ,
\label{eq:BO-Atom} 
\eeqan 
where $|AA \rangle $ and $|A^* A^* \rangle $ is the electron wave
function corresponding to a pair of well separated clusters in their
atomic ground state and excited states respectively and where 
we assume a system with no permanent electric dipole. Driven by
this compelling molecular analogy we want to analyze the NN interaction under
similar dynamical assumptions.

The generalization to the NN system is straightforward, by just
replacing $V_{\rm dip}$ by the One-Pion-Exchange (OPE) potential
$V_{1\pi}$ and was already discussed in Ref.~\cite{RuizArriola:2009vp}
within the context of chiral soliton models and the associated
long-range spin-flavor universality.
One considers the colorless nucleons as two quark clusters which in the chiral quark model exchange a colorless pion at large distances.
The mutual (chiral) polarizability causes attraction between the nucleons, exactly in the same way as for atom-atom interactions and
 equivalently, using Eq.~\eqref{eq:BO-Atom}, one can obtain an optical potential where the effect of excited states as the $\Delta$
 is included perturbatively,
\beqa
V_{2N} (\boldsymbol r) =  V^{1\pi}_{NN,NN} (\boldsymbol r) +
 2\ \frac{|V^{1\pi}_{NN,N\Delta} (\boldsymbol r) |^2}{ M_{N}-M_{\Delta}}
+ \frac{1}{2}\ \frac{|V^{1\pi}_{NN,\Delta\Delta} (\boldsymbol r) |^2}{M_{N}-M_{\Delta}} + {\cal O}(V^3)  \, ,
\label{eq:oppenheimer} 
\eeqan 
where $V^{1\pi}_{NN,NN}$ is the NN OPE potential and $V^{1\pi}_{NN,N\Delta}$ and $V^{1\pi}_{NN,\Delta\Delta}$ are 
OPE transition potentials
\footnote{This potential depends on the coupling constants $f_{\pi NN}$ and $f_{\pi N\Delta}$.
We can use the relation $f_{\pi NN} = g_A m_\pi / (2 f_\pi)$ and the $SU(N_c)$ relation~\cite{Karl:1984cz}
 $f_{\pi N\Delta} / f_{\pi NN} = 3\sqrt{ (N_c - 1) (N_c + 5)} / ( \sqrt{2} (N_c+2) )$ such that the only {\it free} parameter
 is actually $g_A$ having an admissible value in between $1.26$ and $1.29$.
}.
Eq.~\eqref{eq:oppenheimer} reproduces exactly the Skyrme soliton model result of Refs.~\cite{Walet:1992zza,Walet:1992gw}.
At very short distances Eq.~\eqref{eq:oppenheimer} behaves like a vdW potential $\sim -g_A^4/( \Delta f_\pi^4 r^6 )$ and in fact it reduces to 
the Chiral Two-Pion-Exchange (ChTPE) potential at NLO-$\Delta$ ~\cite{Kaiser:1998wa,Krebs:2007rh} with the identification $h_A/g_A = f_{\pi N\Delta}/(2 f_{\pi NN})$. 
Moreover, although both potentials are not completely equivalent 
they are very similar even at intermediate distances which explain why we
achieve results for most of NN observables looking very much like
those of more sophisticate chiral potentials.


\section{Results}

The BO-vdw potential, Eq.~\eqref{eq:oppenheimer}, presents a short
distance singularity and to deal with it we use the method of
renormalization with boundary
conditions~\cite{PavonValderrama:2005wv}.  In
\cite{RuizArriola:2009vp} we showed satisfactory results for $^1S_0$
and $^3S_1-^3D_1$ phase shifts. Here and for the sake of brevity, we
concentrate on deuteron properties.  A more detailed study will be
presented elsewhere.
The deuteron is solved by fixing its binding energy $B_d = 2.224575\ {\rm MeV}$,
the D/S ratio $\eta = 0.0256$ and the $^3S_1$ scattering length $a_{^3S_1} = 5.419 \ {\rm fm}$ from which we obtain the properties,
 $A_S = 0.873 (8) {\rm fm^{-1/2}}$,
 $r_m = 1.945 (14) {\rm fm}$,
 $Q_d = 0.2712 (1){\rm fm^2}$,
 $P_D = 7.3 (1.2)\%$,
 $\langle r^{-1}\rangle = 0.468 (8) $,
 $a_{^3D_1} = 6.56 (5){\rm fm^5}$ ,
 $a_{E_1} = 1.549 (1) {\rm fm^3}$
in the case in which we use the $SU(N_c)$ quark model relation with $N_c = 3$ and,
 $A_S = 0.886 (9) {\rm fm^{-1/2}}$,
 $r_m = 1.973 (18) {\rm fm}$,
 $Q_d = 0.2789 (12){\rm fm^2}$,
 $P_D = 6.8 (1.4)\%$,
 $\langle r^{-1}\rangle = 0.44 (1)$,
 $a_{^3D_1} = 6.471 (9){\rm fm^5}$ ,
 $a_{E_1} = 1.689 (3) {\rm fm^3}$
in the case $N_c \to \infty$. The estimate error corresponds to taking the extreme values $g_A =
1.26$  and $g_A = 1.29$.
The renormalized $^1S_0$ and coupled $^3S_1$, $E_1$ and $^3D_1$ waves were already shown in Ref.~\cite{RuizArriola:2009vp}. 
The deuteron electromagnetic form factors in the IA using our
renormalized wave functions are displayed in Fig.~\ref{fig:GCGMGQ-IA},
which within uncertainties are reproduced rather well.

%
\begin{figure}[htb]
  \begin{center}
    \includegraphics[height=5.cm,width=4.5cm,angle=270]{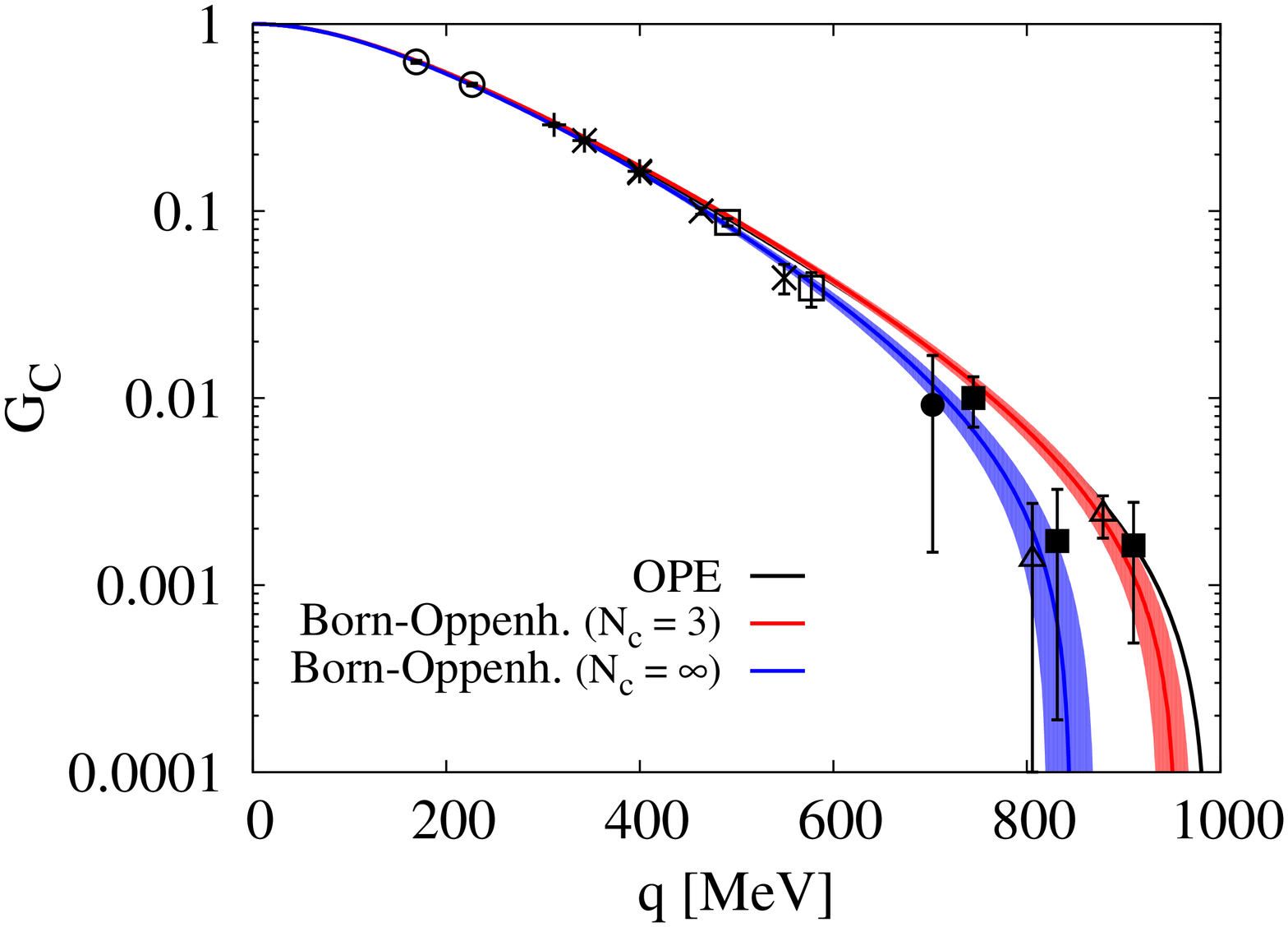}
    \includegraphics[height=5.cm,width=4.5cm,angle=270]{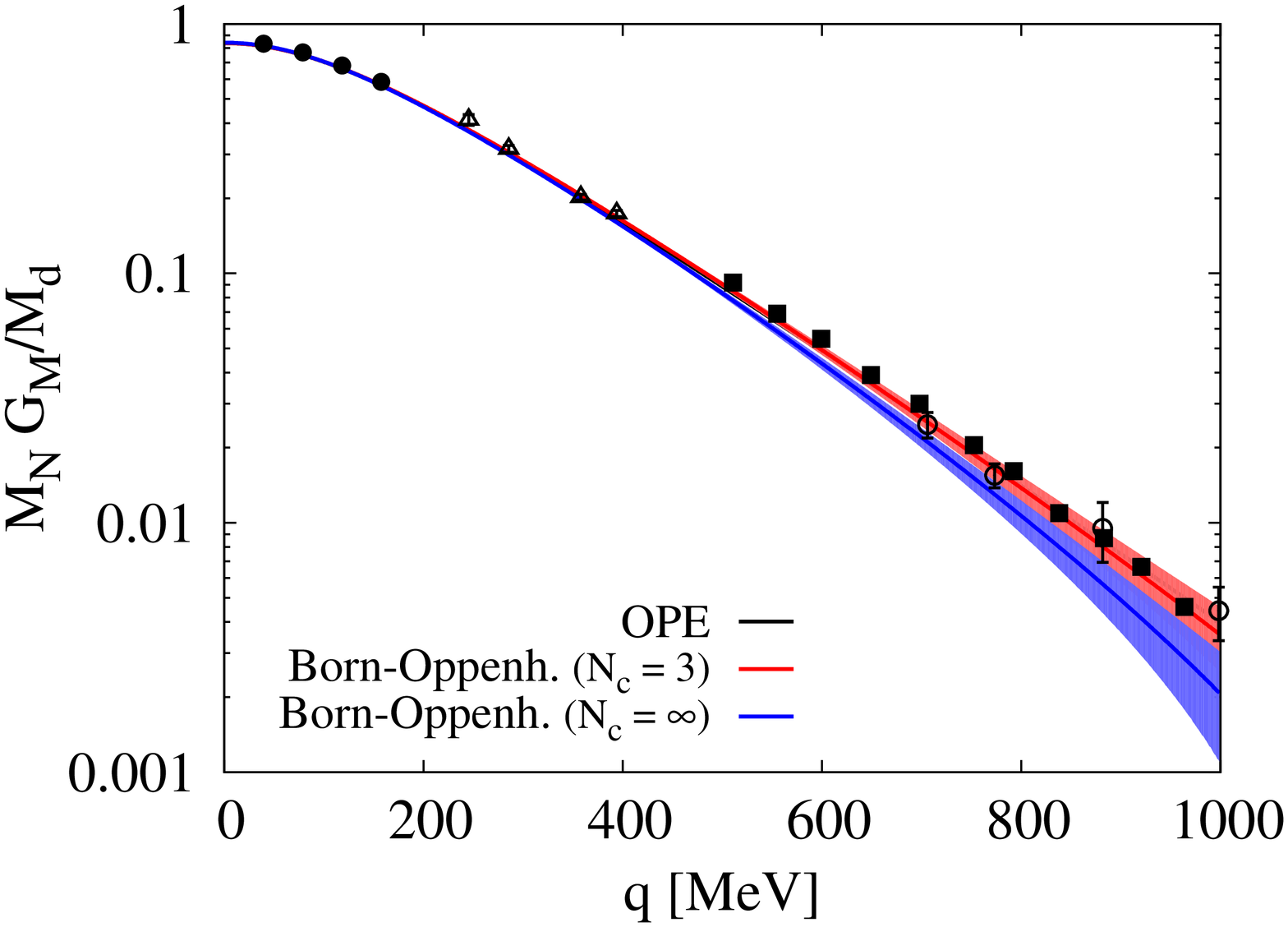}
    \includegraphics[height=5.cm,width=4.5cm,angle=270]{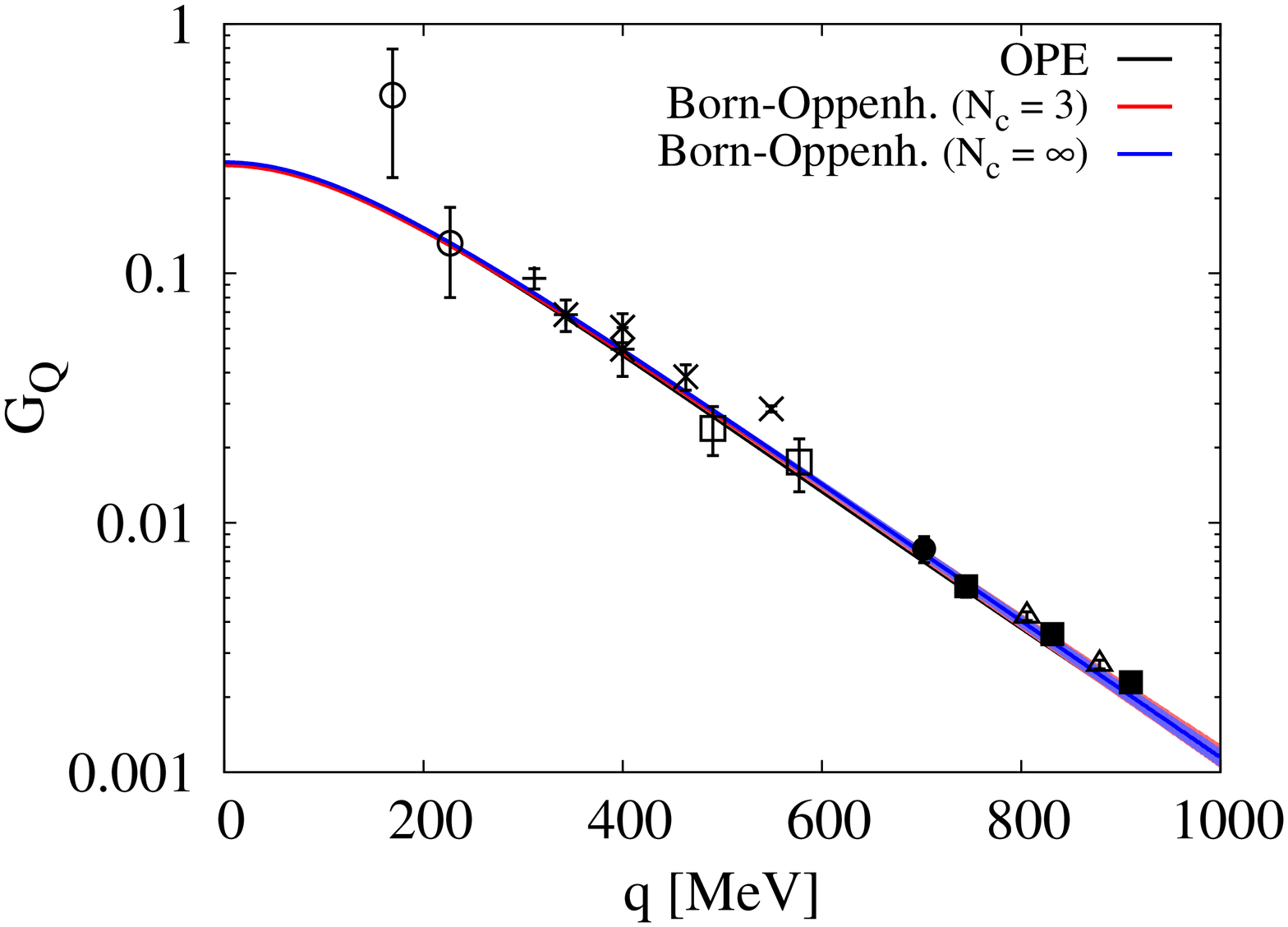}
   \caption{
Deuteron charge $G_C$ (left), magnetic $G_M$ (middle) and quadrupole $G_Q$ (right) form factors in the IA.
%
The dependence with $g_A$ is shown by light bands.
} 
\label{fig:GCGMGQ-IA}.
  \end{center}
\end{figure}
%

%

\section{Three-body force}

The extension to the three nucleon force (3NF) follows from the
generalization of Eq.~\eqref{eq:oppenheimer} for the 3N case, at 2nd
order perturbation theory, being,
\beqa
V_{3N} =
\left\langle NNN \arrowvert V_{OPE} \arrowvert NNN \right\rangle
+ \sum_{NNN\neq HH'H''}
\frac {
\arrowvert \left\langle NNN \arrowvert V_{OPE} \arrowvert HH'H''  \right\rangle \arrowvert^2
} { E_{NNN} - E_{HH'H''} } + {\cal O}(V^3)  \, ,
\label{eq:V3N}
\eeqan
where $H$, $H'$ and $H''$ represent intermediate excited states and $V_{OPE} $ is the sum of
pairwise interactions between nucleons with the exchange of a pion, \ie, 
%
$V_{OPE}(\mathbf r_1, \mathbf r_2, \mathbf r_3) =
V_{OPE}(\mathbf r_{12}) +
V_{OPE}(\mathbf r_{13}) +
V_{OPE}(\mathbf r_{23})
$
and $\mathbf r_{ij} = \mathbf r_i - \mathbf r_j$. Evaluating the matrix elements we obtain,
\beqa
V_{3N} = \sum_{i\neq j} V^{1\pi}_{NN,NN}(\mathbf r_{ij}) 
+  \frac{1}{M_N - M_\Delta} \sum_{i\neq j} \arrowvert V^{1\pi}_{NN,N\Delta}(\mathbf r_{ij}) \arrowvert ^2
+ V^{FM}_{ijk} \, ,
\eeqan
where $ V^{FM}_{ijk}$ is the old Fujita-Miyazawa
3NF~\cite{Fujita:1957zz}.  So, in the BO the 3NF decomposes into a sum
of One-Pion-Exchange Two-Nucleon (1PE-2N) pair interaction,
 a Two-Pion-Exchange Two-Nucleon (2PE-2N) with intermediate $\Delta$ pair
interaction and a genuine Two-Pion-Exchange Three-Nucleon (2PE-3N) interaction.
The emergence of short distance vdW singularities in given channels is evident.
Unfortunately the renormalization of singular three-body problems, even within this
simplified BO approach, has not yet been achieved.
We note that similar interactions have proven to be essential, after
introducing cut-offs, which modify the original interaction below
$2$fm, to describe the binding energies of light nuclei $A \le 8
$~\cite{Pieper:2001ap}.
This suggests that the BO approximation may be
a workable scheme for multi-nucleon forces.

\section{Conclusions}

We have seen how the NN interaction can be faithfully represented as a
vdW force that emerges as in atomic physics where one usually uses
the Born-Oppenheimer approximation. We have calculated the two- and
three-body force at second order in perturbation theory although
higher order may in principle be included. The two nucleon potential
reproduces exactly the Skyrme model result within the same
approximation and its short distances behavior is identical to
ChTPE at NLO-$\Delta$. We have shown results for the deuteron
properties and EM form factors having a very good agreement with
experimental data. In the 3N sector, the BO potential contains the old
Fujita-Miyazawa force as well as a residual 1PE-2N and a 2PE-2N with
$\Delta$.

\acknowledgements{%
ACC thanks Ch. Weiss and R. Schiavilla for discussions. 
Work supported by the Spanish DGI and FEDER funds with grant FIS2008-01143/FIS, Junta de Andaluc{\'\i}a grant FQM225-05.
Authored by a Jefferson Science Associate, LLC under U.S. DOE Contract No. DE-AC05-06OR23177. 
}


%

}  


\end{document}